# Microscopic manipulation of ferroelectric domains in SnSe monolayers at room temperature

*Kai Chang*[1,*,†], *Felix Küster*[1], *Brandon J. Miller*[2], *Jing-Rong Ji*[1], *Jia-Lu Zhang*[1], *Paolo Sessi*[1], *Salvador Barraza-Lopez*[2] *& Stuart S. P. Parkin*[1,*]

1. Max Planck Institute of Microstructure Physics, Weinberg 2, Halle 06120, Germany

2. Department of Physics, University of Arkansas, Fayetteville, Arkansas 72701, USA

ABSTRACT:

Two-dimensional (2D) van der Waals ferroelectrics provide an unprecedented architectural freedom for the creation of artificial multiferroics and non-volatile electronic devices based on vertical and co-planar heterojunctions of 2D ferroic materials. Nevertheless, controlled microscopic manipulation of ferroelectric domains is still rare in monolayer-thick 2D ferroelectrics with in-plane polarization. Here we report the discovery of robust ferroelectricity with a critical temperature close to 400 K in SnSe monolayer plates grown on graphene, and the demonstration of controlled room temperature ferroelectric domain manipulation by applying appropriate bias voltage pulses to the tip of a scanning tunneling microscope (STM). This study shows that STM is a powerful tool for detecting and manipulating the microscopic domain structures in 2D ferroelectric monolayers, which is difficult for conventional approaches such as piezoresponse force microscopy, thus facilitating the hunt for other 2D ferroelectric monolayers with in-plane polarization with important technological applications.

MAIN TEXT:

As the research of ferroelectrics moves towards the 2D limit, the existence of ferroelectricity in monolayer (ML)-thick materials has been confirmed in both van der Waals MLs[1]-[3] and epitaxial perovskites[4]. Compared with traditional perovskite ferroelectric thin films, 2D ferroelectrics bring the freedom of fabricating functional van der Waals heterostructures without the restriction of lattice matching. Many layered ferroelectrics, such as $CuInP_2S_6$[5],[6], $SnTe$[1],[7]-[9], $SnS$[3],[10], $In_2Se_3$[11]-[16], $WTe_2$[17], $MoTe_2$[2], $BA_2PbCl_4$[18][19] and $Bi_2O_2Se$[20] have been discovered in the past few years. In order to gain a deeper understanding of these novel materials, microscopic studies of the ferroelectric domain structures and switching mechanisms become timely tasks. Among common approaches for ferroelectric studies, scanning probe microscopy (SPM) methods are the only category that can simultaneously realize the microscopic characterization and manipulation of ferroelectric domains. Other methods, such as X-ray diffraction, transmission electron microscopy, Raman spectroscopy and second harmonic generation, cannot execute domain manipulation, while devices based on fabricated electrodes can only reveal collective ferroelectric switching behaviors. Currently, piezoresponse force microscopy (PFM) is the most widely applied SPM method for the microscopic studies of 2D ferroelectrics. In PFM experiments, out-of-plane and in-plane polarization components are detected by measuring the deflection and torsion of a cantilever, as responses to the material deformation induced by the vertical electric field applied through a tip fixed at the end of the cantilever[21]. Microscopic imaging and controlled manipulation of the out-of-plane polarization in 2D ferroelectric materials *via* PFM has been demonstrated in $MoTe_2$ MLs[2], freestanding perovskite thin films[22], as well as several-layer thick van der Waals ferroelectrics[5],[11],[13],[15]-[17]. However, although PFM studies of in-plane ferroelectricity has been reported in several-nm thick layered materials[13]-[15],[18]-[20], to the best of our knowledge, none of the current PFM studies of in-plane polarized ferroelectrics has reached the thickness limit of a single van der Waals ML.

As another important SPM method, scanning tunneling microscopy (STM) provides a different viewpoint for the studies of in-plane polarized ferroelectric MLs. Since STM is extremely sensitive to the surface electronic structures of atomically flat samples, in-plane polarization can be detected through measuring the electronic band bending induced by the bound charges at the edges and ferroelectric domain walls of the films. Therefore, the imaging of in-plane polarization in 2D ferroelectric MLs is relatively easy for STM, while the microscopic manipulation of ferroelectric domains turns out to be more difficult than PFM, because the in-plane component of the electric field between the STM tip and the sample is much weaker than the out-of-plane one. Uncontrolled switching through STM tip induced domain motion in the in-plane polarized SnTe MLs at 4.7 K was demonstrated in 2016[1]. Despite of the fact that the first controlled switching experiment of SnS MLs through coplanar electrodes was reported very recently[3], *microscopic manipulation* of the ferroelectric domains in these in-plane polarized 2D ferroelectric MLs is still absent. Furthermore, the ferroelectric transition temperatures of these group-IV monochalcogenide MLs[23]-[28] need to be determined in a reliable way because these values are important for applications. Here we employed a variable temperature STM to controllably switch the in-plane polarization *via* domain manipulation in SnSe MLs at room temperature, and demonstrated its ferroelectric transition temperature to be as high as 380 ~ 400 K, close to that of $BaTiO_3$[29]. These results have demonstrated the ability of STM in characterizing and manipulating these in-plane polarized 2D ferroelectric MLs, and open the door to the potential fabrication of sub-nanometer-thick ultrathin memories[30],[31], non-linear optic[32],[33], spintronic[34]-[36] and valleytronic[37] devices that operate at room temperature based on group-IV monochalcogenide MLs.

We first establish the conformation of ultrathin SnSe nanoplates grown by molecular beam epitaxy (MBE) on graphitized 6H-SiC(0001) surfaces. Belonging to the noncentrosymmetric space group $Pnm2_1$, a SnSe ML contains two strongly bonded atomic layers vertically separated by 2.8 Å. The lattice vectors $a_1$ and $a_2$ are parallel and perpendicular to the in-plane spontaneous polarization $P$, respectively. As seen in Figure 1a, the edges of rectangular SnSe ML plates occur along the ⟨11⟩ directions, with two edges hosting positive bound charges and the other two having negative bound charges. The SnSe plates display three relative crystalline orientations where $a_1$ ($a_2$) aligns with the graphene's zigzag (armchair) direction. Accordingly, their in-plane polarization points along one of the six directions shown in Figure 1b given that a crystalline orientation allows two conformations with antiparallel polarizations. Such highly oriented growth, which is not seen for SnTe MLs on this substrate[1],[8],[38], is further confirmed by reflection high energy electron diffraction (RHEED) patterns and large-scale d$I$/d$V$ mapping images (Figure S1 in Supplementary Information). By reproducing the moiré pattern between the SnSe MLs and graphene through a lattice simulation, we determine the lattice parameters for these SnSe MLs at room temperature to be $a_1 = 4.35 \pm 0.02$ Å and $a_2 = 4.26 \pm 0.02$ Å (Figures S2-S4). The lattice parameter $a_2$ agrees with graphene's periodicity along its armchair direction ($\sqrt{3} \times 2.46$ Å = 4.26 Å), which is the reason for the highly oriented growth of SnSe MLs. For comparison, SnTe MLs has a significant lattice mismatch with graphene ($a_1 = 4.58$ Å and $a_2 = 4.44$ Å, 4.2% mismatch along the armchair direction), thus the relative orientation of SnTe MLs on graphene was found to be random[1].

The spatially resolved d$I$/d$V$ spectra in Figure 1c—approximately proportional to the scanned surface's local density of states (LDOS)[39],[40]—uncovers an in-plane polarization by displaying band bending towards opposite directions at opposite edges of a single-domain SnSe ML plate, distinct from the band bending towards the same directions at opposite edges in non-polar semiconductors. The edges of a SnSe plate are roughly 45° away from the direction of polarization, generating a bound charge density of $\sqrt{2}P/2$ at the edges. The valence band maximum (VBM) lies at −0.44 eV and the conduction band minimum (CBM) at 1.69 eV at the center of this 50 nm wide SnSe ML plate, yielding an electronic band gap of 2.13 eV at 1.8 K. The CBM and VBM show +0.24 eV (upward) and −0.31 eV (downward) band bending at opposite edges in Figure 1c. Fitting these band edge profiles with an exponential function $V(x) = A \exp[-(x - x_0)/L_0] + V_0$, we obtain decay lengths of $L_0 = 4.22 \pm 0.23$ nm for upward bending, and $L_0 = 5.89 \pm 0.36$ nm for downward bending (Figure S5). Such an exponential decay is due to screening from the metallic graphene substrate. Without screening, the decay would take a much slower logarithmic form in a 2D ferroelectric with in-plane polarization, in analogy with the electric potential generated by charged wires. To further confirm that the band bending is consistent with the direction of polarization, we also acquired spatially resolved d$I$/d$V$ spectra along directions parallel and perpendicular to the polarization of a SnSe ML plate (Figs. 1d and e).

As expected, no significant band bending is resolved along the direction perpendicular to the polarization.

Though still resolvable, band edges become less clear on d$I$/d$V$ spectra acquired at room temperature (Figure S6). In that case, and as seen in Figsures 2a and b, topographic and d$I$/d$V$ mapping images still clearly display opposite charge accumulation at opposite edges resulting from band bending. The tunneling current is roughly proportional to the sample's LDOS integrated between $eV_s$ and $E_F$, where $V_s$ is the sample bias voltage and $E_F$ is the Fermi level set at 0 eV. The tip moves away from (closer to) the surface at a place with a higher (lower) integrated LDOS when the STM is operated in a constant current mode (keeping tunneling current $I_t$ unchanged in the scan), generating a higher (lower) apparent height $z$. Therefore, on an atomically flat surface, an edge with upward (downward) band bending shows a higher (lower) apparent height at a negative $V_s$ close to VBM. At $V_s = -0.2$ V and $I_t = 2$ pA, the topographic image of the SnSe ML plate in Figure 2a displays a contrast of 1.0 Å at opposite edges ($z_{up}/z_{down} = 1.2$), while the d$I$/d$V$ image in Figure 2b shows a contrast of 0.6 pS [(d$I$/d$V$)$_{up}$/(d$I$/d$V$)$_{down}$ = 2.3], both confirming an in-plane polarization parallel to $a_1$.

The highly oriented growth prohibits the existence of 90° domain walls that were predominantly seen in SnTe MLs[1]. Only 180° domain walls are observed in SnSe MLs, either in a straight (Figure 2c) or zigzag shape (Figure 2d). An atom-resolved image reveals identical lattice vectors in neighboring domains (Figure S7) in agreement with the 180° domain orientation. The straight wall in Fig. 2c is 12° away from $a_1$, yielding a bound charge density $\lambda_P = 0.42P$, where $P$ is the in-plane polarization estimated to be $1.5 \times 10^{-10}$ C/m from our first principles calculations. The walls in Figure 2d have domains joining "tail-to-tail" and the 34° zigzag angle leads to a bound charge density of $\lambda_P = 0.58P$. Interestingly, the 180° domain walls in Figures 2c and d exhibit high

d$I$/d$V$ intensity comparable to that at the edges with upward band bending, even though the expected bound charge density is lower at the domain walls ($\lambda_P = 0.71P$ at the edges). This is probably because of the existence of a bound state at the VBM of the 180 ° domain walls, according to our first principles calculations (Figure S8). Another intriguing fact is that the 180 ° domain walls are usually mildly negatively charged (tail-to-tail), with most of the angles between the domain wall and the polarization below 30 °. Further studies are needed to unveil the mechanism behind this phenomenon.

Now that the SnSe ML plates are characterized, we discuss the principal contribution of the present work: the controlled ferroelectric domain manipulation at room temperature, including moving, creating and eliminating the 180 ° domains, achieved by applying bias voltage pulses with an STM tip situated at a lateral distance $d_0$ away from the SnSe plate as illustrated in Figure 3a. The width of the uprising edge of these pulses is 0.14 ms (Figure 3b), much longer than the relaxation time of the carriers in graphene[41], thus the electric field induced by the pulse can be regarded as quasi-static. The pulse voltage $V_P$ lasts for $t_P = 50$ ms, unless otherwise specified. In Figures 3c-f, the domain manipulation process is demonstrated as a series of pulses were applied with the STM tip placed at $d_0 = 20$ nm away from a corner of the plate. There were two domains and a 180 ° domain wall across the plate at the onset; the domain with ***P*** parallel to ***E***$_{//}$ expanded and the one with ***P*** antiparallel to ***E***$_{//}$ shrunk as the first $V_P = -5$ V pulse was applied (Figure 3d). This manipulation not only induced a change of domain sizes, but also caused the rotation of the domain wall. Specifically, the angle between the domain wall and the polarization was changed from 28 ° to 7 °, reducing $\lambda_P$ from 0.94$P$ to 0.24$P$, which is consistent with the suppressed d$I$/d$V$ intensity at the domain wall after the manipulation. Such domain wall rotations were common in the domain manipulations thata we performed this way. The whole plate turned into a single domain with ***P*** parallel to ***E***$_{//}$ (Figure 3e) after a second −5 V pulse took place. A new domain was created as a +8 V pulse was applied, and part of the polarization in this plate was switched to the opposite direction (Figure 3f). Finally, another +8 V pulse set the whole plate into a single domain again but inverting ***P*** from its initial direction (Figure 3g). The polarization of a SnSe ML plate can be repeatedly and controllably switched this way, and additional experiments are given in Figure S9. In Supplementary Information, we also present a video to show back-and-forth domain wall motion by the applying bias voltage pulses of opposite signs with the STM tip staying at the same location.

It should be noted that the d*I*/d*V* images acquired after bias voltage pulses were applied are only suitable for inspections of the distribution of ferroelectric domains, while quantitative analysis of band bending profiles should be based on the data acquired right after the STM tip calibration (in our case, on a Au(111) surface), such as in Figures 1c and 2b. This is because the strong electric field during the pulse could modify the tip state and the corresponding tunneling matrix elements, consequently inducing quantitative deviations for the d*I*/d*V* images. The interaction between the STM tip and the ferroelectric MLs during the switching process likely contain rich physics for future studies.

Successful manipulation operations require pulses to be applied on the graphene substrate and away from the SnSe ML plate, because the STM-tip-induced electric field has an out-of-plane component $E_z$ several orders of magnitude larger than its in-plane component $E_{//}$ (Figure 4a). Such a huge $E_z$ can lead to the electric breakdown when the tip is above of a SnSe ML plate. For instance, assuming $V_P$ = 5 V and a tip-sample distance of 5 Å, $E_z$ can easily reach $10^8$ V/cm beneath the STM tip.

Judging from Figure 4b, which shows statistics of many single bias voltage pulse experiments, a pulse with shorter $d_0$ and larger $|V_P|$ has a higher probability of inducing domain wall motion. The uncertainty of the effect of a single pulse might be a result of the complicated shape of the STM tip and the environment of the SnSe ML plates—domain wall pinning could happen due to the defects, wrinkles or atomic steps on the substrate. In order to quantitatively determine the critical field of domain wall motion, $E_{//,c}$, we carried out a series of experiments in which $|V_P|$ was gradually increased until domain wall motion was observed (see Figure S8 for details). Comparing the experimental critical pulse voltages $|V_{P,c}|$ with the $E_{//}$ obtained from numerical simulations in Figure 4c (see also Figure S10 for details of the simulations), we derived the corresponding critical fields for different $d_0$ in Figure 4d. Our data suggests that for $d_0 \geq 30$ nm, $E_{//,c}$ converges to $1.4 \pm 0.2 \times 10^5$ V/cm, which we consider to be the intrinsic critical field for domain wall motion. The $E_{//,c}$ for shorter $d_0$ is higher, probably because of nonlinear electrostrictive effects induced by the very large $E_z$ at these distances.

We now determine the transition temperature $T_c$ above which the spontaneous polarization is suppressed. In Figure 5, we used a variable temperature STM to study the temperature dependence of band bending at the edges of the SnSe ML plate shown in Figure 2a. The contrast of both apparent heights (Figures 5a-f) and d$I$/d$V$ (Figures 5g-l) at opposite edges—both depending on the magnitude of band bending—decreases as the temperature is increased and becomes totally indistinguishable at 400 K (Figures 5e,k,q). Band bending reappears as the temperature is decreased to 308 K, regaining the magnitudes registered before heating (compare Figures 5a,g,m to Figures 5f,l,r). This implies that the $T_c$ of SnSe MLs on graphene is between 380 K and 400 K, similar to that of bulk BaTiO$_3$ (396 K)[29], a well-known perovskite ferroelectric. For comparison, SnTe MLs on graphene substrates have a $T_c = 270$ K [1], a value below room temperature that may hinder practical applications. The larger value of $T_c$ obtained for SnSe MLs on graphene is consistent with a larger energy required to turn the ferroelectric phase into the paraelectric phase as the chalcogen atom (Se versus Te) becomes lighter[23],[42].

In the end, despite the preferential orientation during growth discussed previously, SnSe ML plates can indeed be controllably moved by the STM tip on the graphene substrate without observable damage (Figure S11), still implying a weak interaction between SnSe and the substrate. This makes the creation of heterostructures containing SnSe MLs by transferring and stacking techniques, or even by *in situ* scanning probe manipulations, likely to be feasible.

This ferroelectric domain characterization and manipulation method by STM is a useful complementary technique for the traditional coplanar electrode experiments on in-plane polarized 2D ferroelectrics, such as the recently reported ferroelectric switching of SnS MLs[3]. On the one hand, measuring the collective ferroelectric behaviors, coplanar electrode experiments are more defined and quantitatively precise; on the other hand, STM has unique ability of microscopically imaging and manipulating the find domain structures in these ferroelectric MLs. The combination of these two approaches should be beneficial for future comprehensive studies.

The discovery that highly oriented SnSe MLs on graphene are 2D ferroelectrics whose ferroelectric domains can be controllably manipulated at room temperature brings predicted effects and device concepts based on group-IV monochalcogenide MLs—such as linearly-polarized-light-controlled valley selective excitations[24],[43],[37], shift current photovoltaics[32], intrinsic valley Hall effects[37], in-plane ferroelectric tunneling junctions[30],[31], and nonlinear photocurrent switching devices[33]—one step closer to reality. Additionally, we envision that the coupling between existing 2D ferromagnets and 2D ferroelectrics via vertical or horizontal heterojunctions may lead to an eventual deployment of layered artificial multiferroics.

**Methods**. *Sample preparation:* SnSe monolayer plates were grown on graphene substrates via van der Waals molecular beam epitaxy in an ultra-high vacuum (UHV) chamber with a base pressure of $1 \times 10^{-10}$ mbar. Graphene substrates were prepared following a sequential direct current annealing procedure of nitrogen doped 6H-SiC(0001) (Si face). The substrate temperature was monitored by a high-precision infrared pyrometer. The 2 mm × 10 mm sized SiC substrate was first degassed at 500 °C overnight, then annealed at 900 °C in Si flux for 15 min to form a Si-rich 3 × 3 reconstruction. Finally, the substrate was annealed at 1400 °C for 10 min to graphitize the surface. SnSe molecules were evaporated from a home-built thermal evaporator with an h-BN crucible containing 99.999% SnSe granules from Alfa Aesar. During growth, the SnSe evaporator was kept at ~450 °C and the substrate temperature varied from 70 °C to 220 °C, depending on the specific plate shape desired. During growth, the substrate was heated by radiation from tungsten filaments, and temperature was read from a thermocouple. Higher substrate temperatures generate thicker plates with uniform thicknesses and straight edges, while plates grown at lower substrate temperatures are more irregular and have more ~6 Å high steps at their surfaces. Rectangular SnSe monolayer plates were grown in a two-step process: (i) by initially depositing irregularly shaped monolayer plates with a coverage lower than 0.05 monolayers at a substrate temperature of 70 °C and (ii) a subsequent annealing up to 240 °C for 1 h to turn the plates rectangular. If larger monolayer plates are desired, one can deposit SnSe again at the substrate temperature of 240 °C, at which SnSe plates tend to grow horizontally.

*Variable temperature scanning tunneling microscopy (VT-STM):* As-grown samples were transferred into an Omicron VT-STM-XT system connected to the growth chamber without leaving the UHV environment. Mechanically sharpened Pt/Ir (80/20) alloy tips calibrated on an Au(111) standard sample were used for scanning. For the measurement of d$I$/d$V$ spectra and mappings, a sinusoidal modulation voltage of 30 mV with a 713 Hz frequency was added to the bias voltage. In the ferroelectric switching experiments, the tip was first stabilized at the parameters for scanning (typically −0.2 ~ −0.4 V, 2 pA), and then the feedback loop was turned off when a pulse voltage (typically ±3 ~ ±8 V) was applied. The feedback loop was turned on immediately at the end of a pulse. A Lakeshore 335 temperature controller was used for the variable temperature experiments. The temperature ramping speed was limited to <5 ℃/min, and the STM kept scanning during the ramping process in order to track the thermal drift of the sample. At each temperature setpoint, the temperature was stabilized for at least 30 min to reduce thermal drift. For collecting the d$I$/d$V$ spectra at room temperature, the bias voltage was scanned both forward and backward to compensate the drifting of tunneling junction width. Acquiring each spectrum took only 3~4 seconds. 20 spectra were averaged at each point in order to reduce the noise.

*Low temperature scanning tunneling microscopy (LT-STM):* Low-temperature measurements were performed using a cryostat (Oxford Instruments) equipped with an UHV insert hosting the STM head (Sigma Surface Science). A steady state temperature of 1.8 K is achieved by continuously pumping the $^4$He cooling circuit. Samples were transferred from the MBE system to the LT-STM using a vacuum suitcase with a pressure in the $10^{-10}$ mbar range, thus always preserving UHV conditions. STM data were obtained using electrochemically etched W tips. Before measurements, the tips were conditioned on an Ag(111) single crystal. d$I$/d$V$ spectra were acquired by a lock-in technique, using a bias voltage sinusoidal modulation of 30 mV at a frequency of 736 Hz. To avoid any drift, line spectra across SnSe monolayer plates were taken after scanning the very same area for several hours. Each line consists of 200 points and was measured in approximately 6 hours. Positive and negative energy ranges were acquired separately.

*Ab initio calculations:* Ab initio calculations with the *VASP* code[44],[45] that employ the self-consistent van der Waals exchange correlation functional as implemented by Hamada were carried out. In these calculations, 17 unit cells of the SnSe monolayer were almost commensurate with 30 rectangular graphitic cells (each containing 8 atoms) down to less than 0.5% strain. In addition, periodic 180º domains 15 nm wide were created using methods similar to those employed in References [23], [26] and [46], by the vertical stacking of 35 rectangular unit cells with polarization parallel to lattice vector $a_1$, and 35 additional unit cells with polarization antiparallel to $a_1$. The antiparallel unit cells were displaced horizontally until a minimum energy was found, and a subsequent structural optimization employing 25 k-points along the horizontal direction until forces were smaller than 0.01 eV/Å. The projected density of states was produced employing 100 k-points along the horizontal direction, and spin orbit coupling was included in these calculations.

*Numerical simulations of electric field:* The numerical simulations were performed with the COMSOL software. The electric field was set as static in the simulations. The diagram and parameters of the model and the calculated results are included in Figure S9 of Supplementary Information.

FIGURES

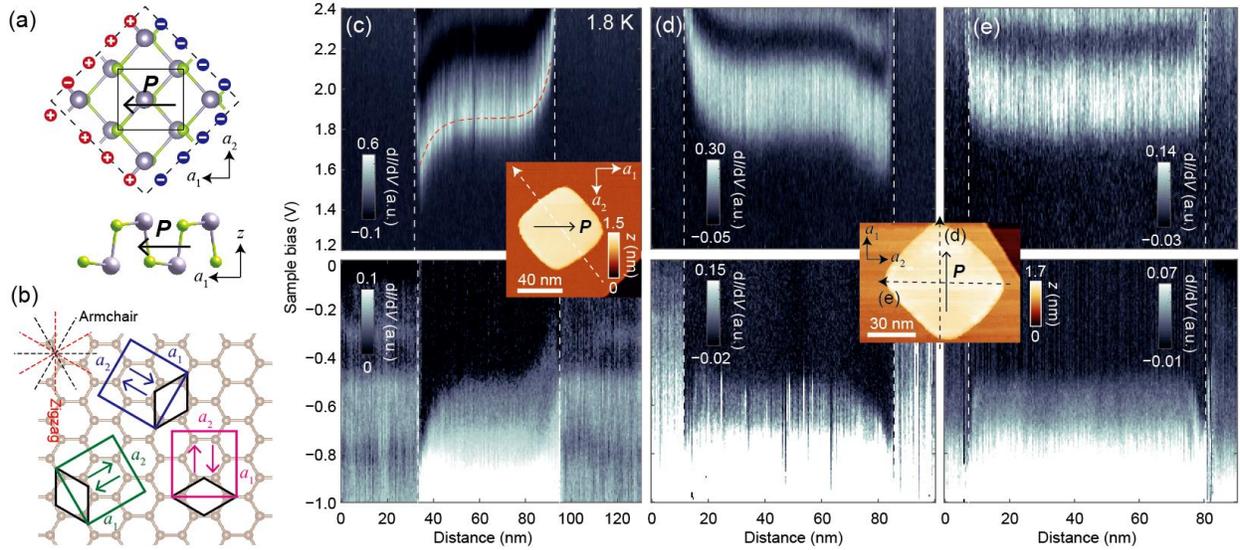

**Figure 1.** In-plane spontaneous polarization in SnSe monolayer plates. (a) Lattice structure of SnSe monolayer. The solid rectangle indicates a unit cell. Dashed lines indicate the preferred edge orientations of these plates. The signs of bound charges at opposite edges are labeled. (b) Schematic diagram of three possible crystalline orientations of SnSe monolayers. The rhomboids and rectangles represent the unit cells of graphene and SnSe, respectively. Arrows in the rectangles indicate the directions of polarization allowed in this configuration. (c) Spatially resolved $dI/dV$ spectra along the dashed arrow across the SnSe monolayer plate shown in the inset, obtained at 1.8 K. Setpoints: $V_s = 3.0$ V, $I_t = 50$ pA for positive $V_s$; $V_s = -1.0$ V, $I_t = 50$ pA for negative $V_s$. The red dashed curves are exponential fittings of the band bending profiles. (d),(e) Spatially resolved $dI/dV$ spectra acquired from the same SnSe monolayer plate, along the directions parallel (d) and perpendicular (e) to the polarization, as the inset shows. The conditions of measurements are the same as in (c).

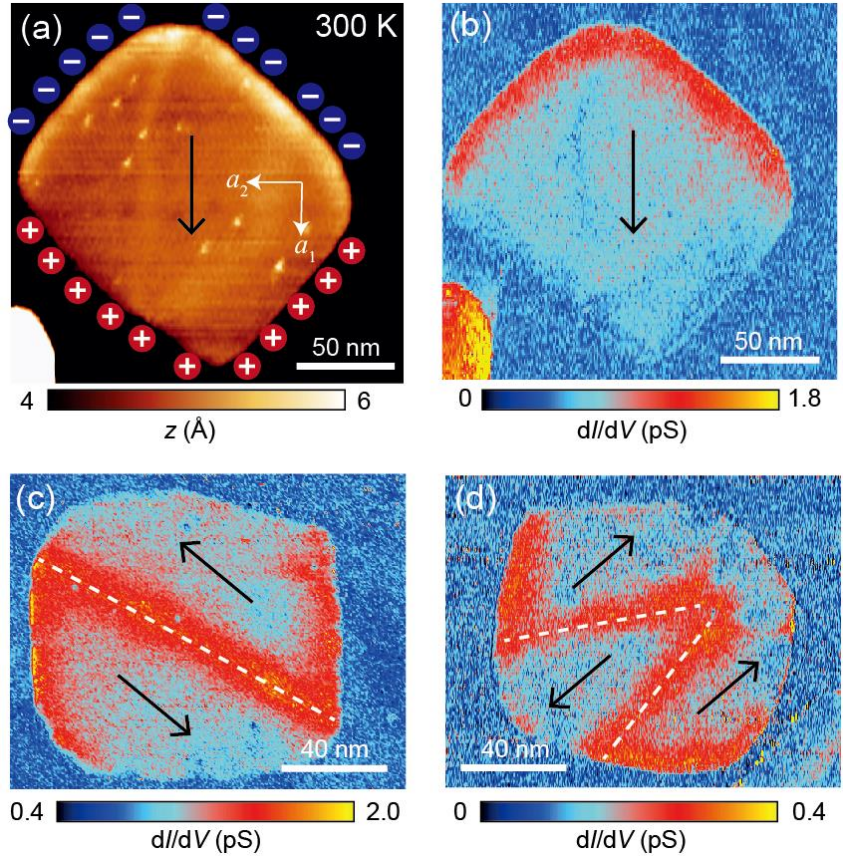

**Figure 2.** Ferroelectric domains in SnSe monolayer plates at room temperature. (a),(b) Room temperature topography and simultaneously recorded d$I$/d$V$ images of a SnSe monolayer plate. Setpoint: $V_s = -0.2$ V, $I_t = 2$ pA. (c),(d) d$I$/d$V$ images of SnSe monolayer plates with 180 ° straight (c) and zigzag (d) domain walls. The domain walls are indicated by the white dashed lines. Setpoints: $V_s = -0.2$ V, $I_t = 2$ pA for (c), $V_s = -0.35$ V, $I_t = 2$ pA for (d).

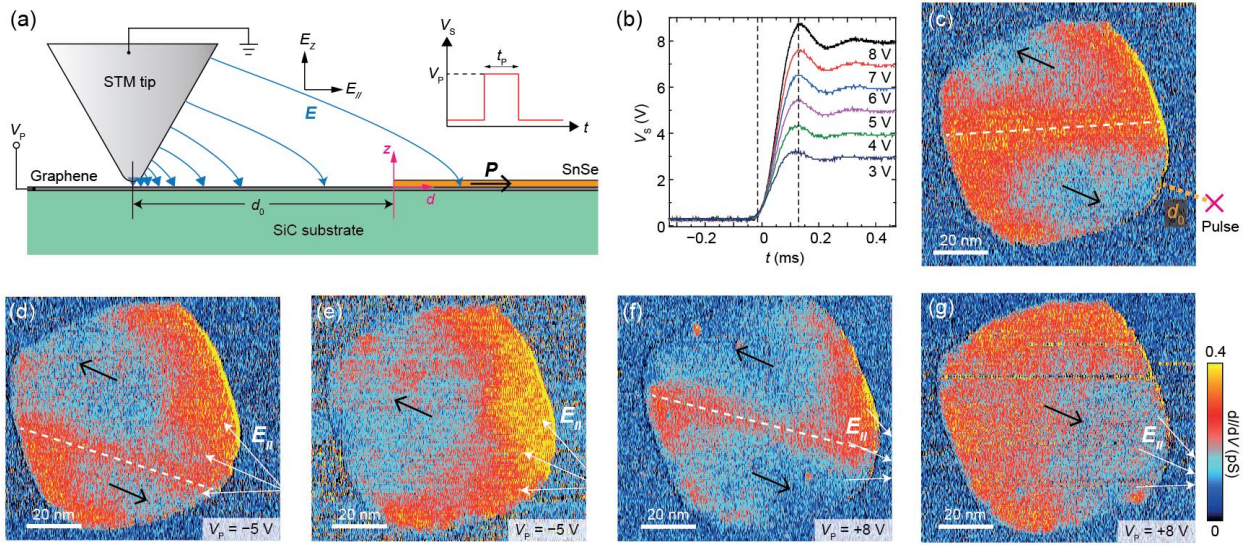

**Figure 3.** Controllable ferroelectric switching of SnSe monolayer plates. (a) Schematic of ferroelectric switching achieved by applying a bias voltage pulse $V_P$ at a point on the graphene substrate close to a SnSe monolayer plate. The corner of SnSe closest to the STM tip was set as $d = 0$, and the upper surface of graphene was set as $z = 0$. (b) Rising edges of the bias voltage pulses. Dashed lines indicate the onset and maximum values of the bias voltage. (c)-(g) Consecutive d$I$/d$V$ images of a ferroelectric switching sequence in a SnSe monolayer plate. Setpoints: $V_s = -0.35$ V, $I_t = 2$ pA. The pulses were applied at the same point indicated in (c). The widths of all the pulses were 50 ms. The directions of the in-plane components of tip-induced electric fields are indicated by the white arrows. All the data in this figure were collected at room temperature.

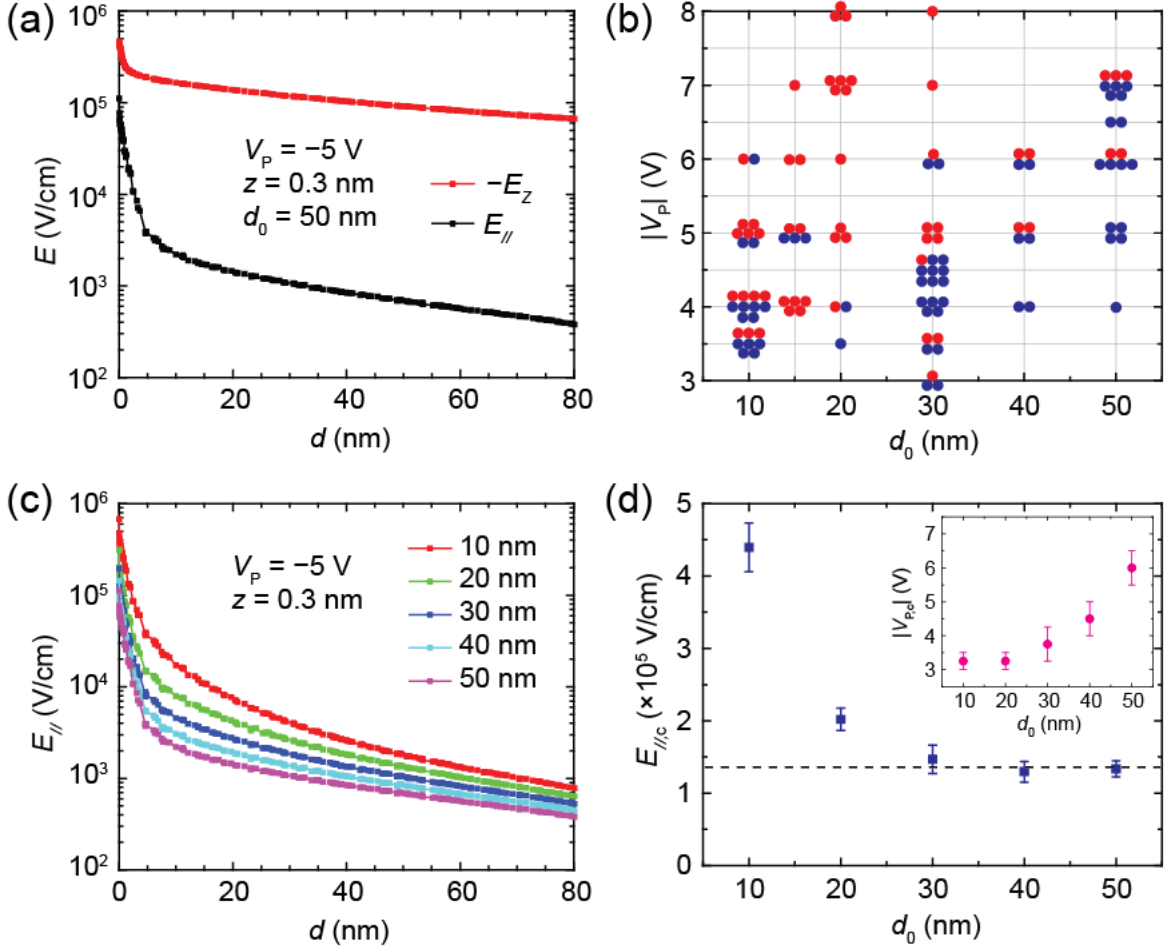

**Figure 4.** Quantitative studies of the domain wall motion induced by bias voltage pulses. (a) Simulated out-of-plane and in-plane electric fields along a horizontal line along the $d$ axis (see Fig. 2(a)) at the height of $z = 0.3$ nm. (b) Statistics of a series of bias voltage pulse experiments with different $V_P$ and $d_0$. Each dot represents a single pulse applied. Those pulses that successfully moved a domain wall are shown in red, and those that did not induce domain wall motion are in blue. (c) Simulated in-plane electric fields with $V_P$ fixed while varying $d_0$. (d) The experimental $d_0$ dependence of the critical pulse voltages $|V_{P,c}|$ of domain wall motion (inset) and the corresponding $E_{//,c}$ at the closest corner of SnSe.

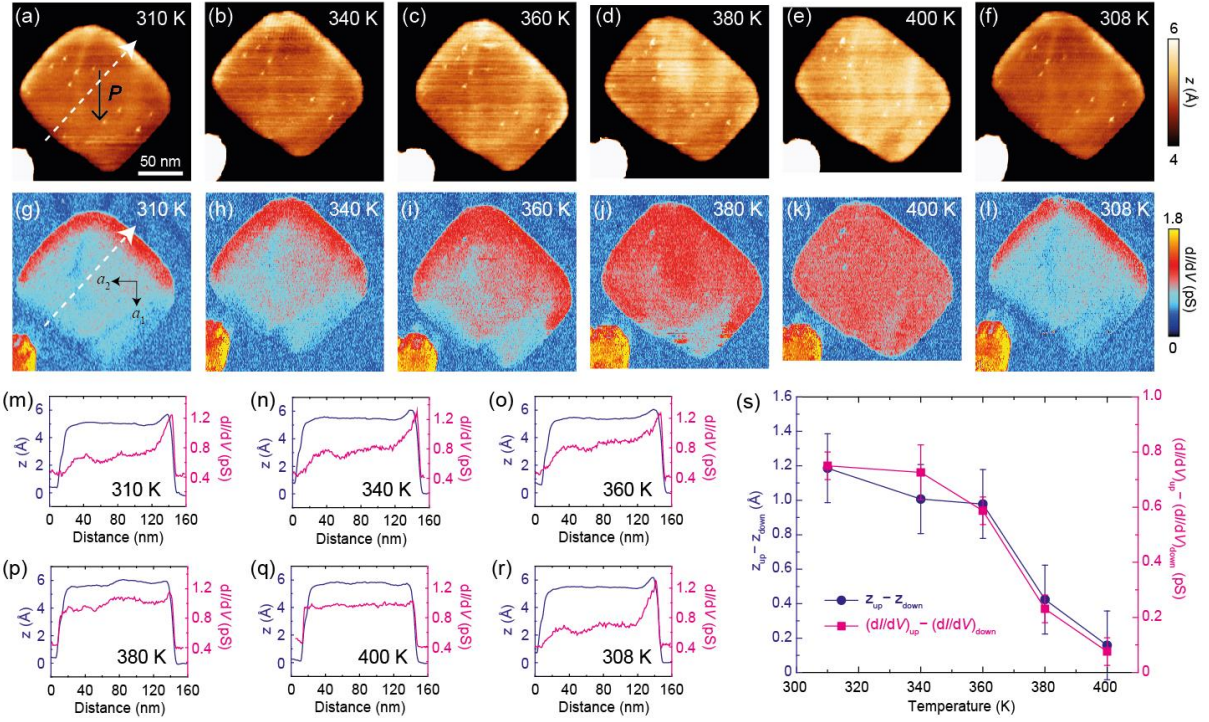

**Figure 5.** Temperature dependence of ferroelectricity in a SnSe monolayer plate. (a)-(f) Topography images of a SnSe monolayer plate at temperatures $T$ increasing from 310 K to 400 K from (a) to (e), and decreasing to 308 K in (f). Setpoints: $V_s = -0.2$ V, $I_t = 2$ pA. (g)-(l) Simultaneously recorded d$I$/d$V$ images corresponding to (a)-(f). (m)-(r), Apparent height and d$I$/d$V$ profiles along the dashed arrows in (a) and (g), extracted from (a)-(f) and (g)-(l), respectively. (s) Evolution of the difference of apparent height and d$I$/d$V$ between the two opposite edges with upward and downward band bending directions, extracted from (m)-(q).

## ASSOCIATED CONTENT

**Supporting Information**.

The following files are available free of charge.

Additional data containing Figures S1-S11. (PDF)

A video showing the consecutive motion of a domain wall induced by a series of pulses. (MP4)

## AUTHOR INFORMATION

**Corresponding Author**

* E-mail: changkai@baqis.ac.cn (K.C.)

* E-mail: stuart.parkin@mpi-halle.mpg.de (S.S.P.P.)

**Present Addresses**

† (K.C.) Beijing Academy of Quantum Information Sciences, Beijing 100193, China

**ORCID**

Kai Chang: 0000-0002-4965-4537

Salvador Barraza-Lopez: 0000-0002-4301-3317

Stuart S. P. Parkin: 0000-0003-4702-6139

**Author Contributions**

K.C. and S.S.P.P. designed the experiments. K.C. and J.-R.J. prepared the samples and performed the VT-STM experiments. F.K. and P.S. conducted the LT-STM experiments. B.J.M. and S.B.-L. carried out the *ab initio* calculations. J.-L.Z. performed the numerical simulations of electric fields. K.C., P.S., S.B.-L. and S.S.P.P. wrote the manuscript. All coauthors read and commented on the manuscript.

**Funding Sources**

Deutsche Forschungsgemeinschaft (DFG, German Research Foundation): PA 1812/2-1

U.S. Department of Energy, Office of Basic Energy Sciences: DE-SC0016139

U.S. National Science Foundation: 0722625, 0959124, 0963249, and 0918970

Arkansas Economic Development Commission

Office of the Vice Provost for Research and Innovation

U.S. DOE Office of Science: DE-AC02-05CH11231

**Notes**

The authors declare no competing financial interest.

ACKNOWLEDGMENT

We thank Z. K. Liu for providing the SiC substrates, and J. D. Villanova, S. P. Poudel and Y. Zhuang for technical assistance. K.C., F.K., J.-R.J., P.S. and S.S.P.P. were supported by Deutsche Forschungsgemeinschaft (DFG, German Research Foundation) – Project number PA 1812/2-1. B.J.M. and S.B.L. were funded by an Early Career Grant from the U.S. Department of Energy, Office of Basic Energy Sciences (Award DE-SC0016139). Calculations were performed at University of Arkansas' Trestles, funded by the U.S. National Science Foundation (Grants 0722625, 0959124, 0963249, and 0918970), a grant from the Arkansas Economic Development Commission, and the Office of the Vice Provost for Research and Innovation, and at Cori at NERSC, a U.S. DOE Office of Science User Facility operated under Contract No. DE-AC02-05CH11231.

ABBREVIATIONS

MBE, molecular beam epitaxy; STM, scanning tunneling microscopy; 2D, two-dimensional; ML, monolayer; PFM, piezoresponse force microscopy; LDOS, local density of states; RHEED, reflective high energy electron diffraction; VBM, valence band maximum; CBM, conduction band minimum.

BRIEFS

Controlled manipulation of the ferroelectric domain walls at room temperature via a scanning tunneling microscope has been demonstrated in monolayer SnSe, whose ferroelectric transition temperature is found to be close to 400 K.

TOC FIGURE

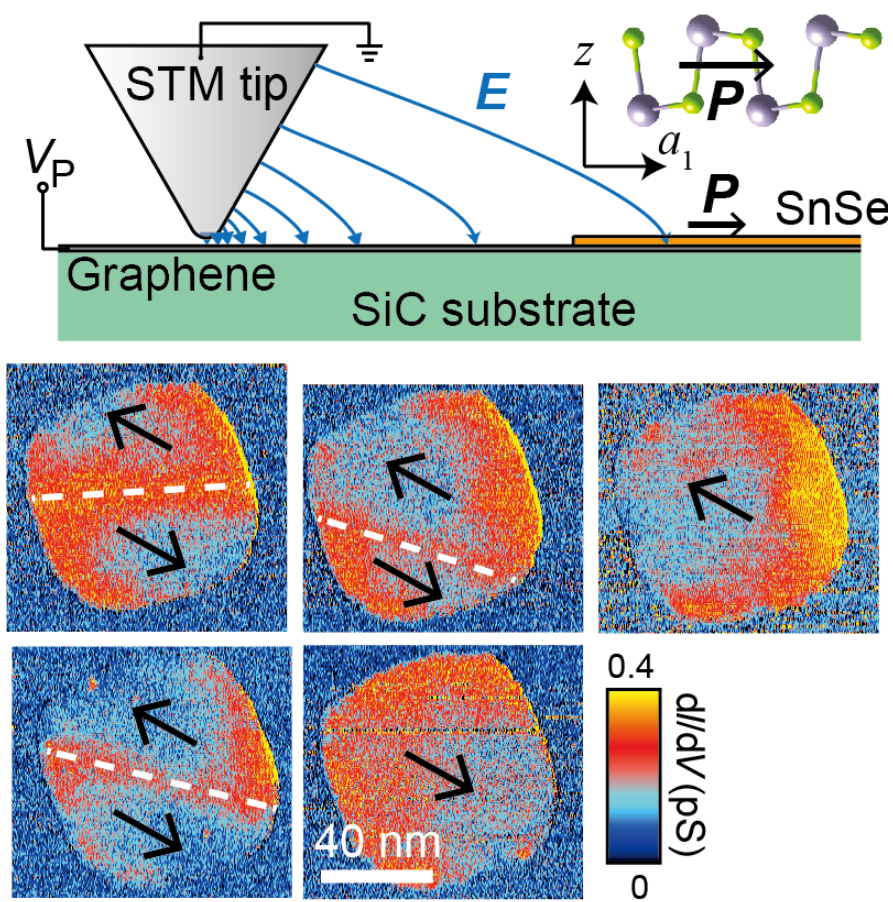